\documentclass{article}

\usepackage{PRIMEarxiv}
\RequirePackage{fix-cm}
\usepackage[utf8]{inputenc} 

\usepackage{fancyhdr}       
\usepackage{graphicx}       

\usepackage{biblatex}
\usepackage{amssymb} 
\usepackage{amsmath} 
\usepackage{amsthm} 
\usepackage{mathtools}    
\usepackage{enumerate}
\graphicspath{ {./images/} }
\newtheorem{proposition}{Proposition}
\newtheorem{lemma}{Lemma}
\newtheorem{definition}{Definition}
\newtheorem{note}{Note}
\title{Mathematical representation of the structure of neuron-glia networks
}

\author{
  Marco Pe\~{n}a-Garcia$^{*}$, Francesco Pe\~{n}a-Garcia, Walter Cabrera-Febola \\
  Group of Natural Structures and Theoretical Research \\
  Universidad Nacional Mayor de San Marcos \\
  Lima, Per{\'u}\\
  \texttt{$^{*}$marco.andre.626@gmail.com} \\
   \And
  Nelson Castro \\
  Department of Model Risk Management \\
  Banco de Cr{\'e}dito del Per{\'u} \\
  Lima, Per{\'u}\\
}

\begin{document}
\maketitle

\begin{abstract}
Network representations of the nervous system have been useful for the understanding of brain phenomena such as perception, motor coordination, and memory. Although brains are composed of both neurons and glial cells, neuron-glial networks have been little studied so far. Given the emergent role of glial cells in information transmission in the brain, we developed a mathematical representation for neuron-glial networks ($\Upsilon$-graph). We also defined the concepts of isomorphisms, unnested form (multidigraph) and matrix equation for $\Upsilon$-graphs. Although we found several network motives where the isomorphism between unnested forms does not guarantees the isomorphism between their respective $\Upsilon$-graphs, we found that if the matrix equations satisfy some conditions, the unnested forms isomorphism guarantees the isomorphism between $\Upsilon$-graphs. Finally, we introduced a novel approach to modeling the network shape. Our work presents a mathematical framework for working with neuron-glia networks.
\end{abstract}




\maketitle

\section{Introduction}

Traditionally, it was commonly accepted that only neurons are involved in the brain information processing while glial cells are relegated to support and protect them. A more updated view suggests that glial cells are also involved in the information processing. In fact, one type of glia cell - astrocytes - couples to single synapses between neurons, forming tripartite synapses \cite{gonzalez2019gliotransmission}. Astrocytes have an active role at tripartite synapses. On one hand, they can directly readout signals from the presynaptic neuron (e.g. they express receptors for GABA, dopamine, glutamate, and glycine)\cite{jennings2017dopamine, verkhratsky2020nmda}. On the other hand, they can modulate the activity of the postsynaptic neuron via the release of neuroactive substances called gliotransmitters \cite{gonzalez2019gliotransmission, scimemi2019role}. 

Neuron-driven activation of astrocytes induces an increase in cytosolic calcium  \cite{jennings2017dopamine, verkhratsky2012calcium}. This increase in calcium, in turn, triggers the release of gliotransmitters \cite{covelo2018neuronal}. More importantly, local increases in calcium spreads through the astrocyte syncytium, i.e. astrocyte network, mainly via IP3 diffusion \cite{fujii2017astrocyte,semyanov2019spatiotemporal}. Interestingly, recent studies suggest that the astrocyte network plays a role in neural network phenomena such as -burst synchronization, bursting behavior, barrage firing, and gamma waves- and in cognitive capabilities (e.g. recognition memory) \cite{chever2016astroglial, deemyad2018astrocytes, lee2014astrocytes}. Therefore, studying neuron-glia networks could be critical to understand the brain dynamics of large populations of neurons.  

Although neuron-glia networks have been modelled before, finding that astrocytes modulate dynamic coordination, facilitate ultra-slow oscillation, and increase the occurrence of bursting-like spikes, they lack a formal network representation \cite{chan2017role,tang2013information,wade2011bidirectional}. Neural networks are typically represented using graphs or digraphs \cite{bassett2017network, reimann2017cliques, sizemore2019importance}. These representations allow the study of features such as connectivity, modularity, and centrality. \cite{bassett2017network}. So far, graph-like structures has guided the study of connectomes at all scales \cite{bassett2017network,karwowski2019application, swanson2016cajal}. However, neuron-glia networks have tripartite synapses that relate three cells at once, thus they can not be represented by edges of a graph nor a digraph. 

In this paper, we addressed this problem by defining a representations of neuron-glia networks. In order to achieve this goal, first, we defined a synaptic configuration as a set of ordered pairs that represent the information flux between cells in a synapse. Then, we represented a collection of synapses as a $\Upsilon$-family that relates synaptic configurations to indexes (single synapses), in order to distinguish synapses that share a synaptic configuration. To simplify this representation, we defined $\Upsilon$-multisets that contain identical copies of repeated synaptic configuration. A set of neuron-like cells (neurons and astrocytes) together with an $\Upsilon$-multiset form an $\Upsilon$-graph which represent a neuron-glia network. This representation allowed us to adapt the concept of multidigraph isomorphism to neuron-glia networks. Then, we defined the $\phi$ function that maps $\Upsilon$-graphs with their unnested forms (multidigraphs) which contain only ordered pairs. Using this function, we found network motifs (propositions \ref{prop_case1}-\ref{prop_case4}) where the isomorphism between unnested forms does not guarantee the isomorphism between $\Upsilon$-graphs, meaning misrepresentation by the multidigraphs. However, we also found a mathematical condition (proposition \ref{secondmainprop}) that allows the unnested forms isomorphism to guarantee the $\Upsilon$-graphs isomorphisms. Finally, we introduced a novel approach to model neuron-glia network shape, using $\Upsilon$-families, natural structure theory, and topology theory \cite{cabrera2004natural2,cabrera2004natural1}. Future research could use our definition of $\Upsilon$-graph isomorphism as part of an algorithm that find neuron-glia network motifs in a connectome that includes astrocytes such as the connectome reconstructed by Sizis et al., 2021 \cite{zisis2021digital} (see \cite{lin2024network} for examples of motif analysis of a connectome). Another research avenue would be extending graph measures (e.g. centrality, modularity, and distance) to $\Upsilon$-graphs, in order to analyze connectomics data (see \cite{lin2024network} for examples of use of network metrics to study a  connectome) or to study how these measures correlates with different predicted networks dynamics in dynamical models. 

\section{Basic definitions}

We follow the definitions and notations of \cite{girish2012multiset} and  \cite{girish2009relations} about multisets (msets) and related objects. We also follow the concepts of \cite{bang2008digraphs} about digraphs and related objects. In the following formal definitions, we summarize the concepts about multisets and graph theory that are necessary to understand our theoretical framework. 

\begin{definition}
    \cite{girish2009relations} A collection of elements containing duplicates is called a multiset. Formally, if $X$ is a set of elements, a multiset $M$ drawn from the set $X$ is represented by a function $Count_M$ defined as $Count_M: X \to \mathbb{N}_0$ where $\mathbb{N}_0$ represents the set of non negative integers.
\end{definition}

$Count_M$ associates each element of the set $X$ with its occurrence in the multiset $M$ 
\cite{girish2009relations,girish2012multiset}. The support set of $M$ is denoted by $M^*$ and it is defined as the subset of $X$ in which the occurrence of all its element in $M$ is greater than 0 \cite{girish2012multiset}.
\begin{definition}
    Let two digraphs $D = (V(D),A(D))$ and $H = (V(D),A(D))$ where $V()$ and $A()$ denote the set of vertices and arcs of the digraph, respectively. A digraph homomorphism is a function $f:V(D) \to V(H)$ if it preserves arcs, meaning that $(x,y) \in A(D)$ implies $(f(x),f(y)) \in A(H)$.
\end{definition}

The following definition formalizes the concept of directed pseudograph \cite{bang2008digraphs} to create our multidigraph definition, using multiset theory \cite{girish2009relations,girish2012multiset}. 


\begin{definition}

A finite multidigraph $Q$ is a tuple $(V(Q),A(Q))$ where $V(Q)$ is a finite vertex set and $A(Q)$ is the finite arcs multiset with support set $A(Q)^* \subset V(Q) \times V(Q) $.     
\end{definition}
Note that if all edges appear once, $A(Q)$ is equivalent to a set, and the multidigraph $Q$ is equivalent to a digraph (loop allowed). Note that $A(Q)^*$ is a binary relation on $V(Q)$. We also formalize the multidigraphs homomorphism in the context of multiset theory as follows. 

\begin{definition}
\label{multidig-homo}
Given two multidigraphs $Q$ and $P$ with vertex sets $V(Q)$ and $V(P)$, respectively, an homomorphism is a bijective function $f: V(Q) \to V(P)$ such that for any $(u,v) \in A(Q)^*$ the following is true.
\[ \text{Count}_{A(Q)}((u,v)) = \text{Count}_{A(P)}((f(u),f(v)))\]
\end{definition}
 
If $Q$ is homomorphic to $P$, and $P$ is homomorphic to $Q$, $Q$ and $P$ are isomorphic ($Q \cong P$). 

\section{Theoretical framework} 

Like neurons, astrocytes receive and release neurotransmitters, so we will refer to both as neuron-like cells. In the following subsection, $N$ and $A$ will represent a set of neuron and a set of astrocytes, respectively. Both $N$ and $A$ are sets of neuron-like cells (NL-sets). Additionally, we can define a NL-set $X$ as the union of $N$ and $A$ ($X = N \cup A $). Since astrocyte and neurons are different cellular types \cite{gonzalez2019gliotransmission}, a set of neurons and a set of astrocytes are always disjoint sets (i.e., $N \cap A = \varnothing$).

\subsection{Synaptic configurations}
Synapses allow the information flux from one neuron-like cell to another. In particular, tripartite synapses allow a directed information flux from the pre-synaptic neuron to the post-synaptic neuron, from the pre-synaptic neuron to an astrocyte, and from the astrocyte to the post-synaptic neuron. We can represent this synaptic configuration as a set of ordered pairs (binary relation) where each ordered pair represents a direction of information flux. 

\begin{definition}
\label{def_tri}
A tripartite configuration is represented by $s_t$ that is set containing three ordered pairs. The ordered pairs represent the information flow of a tripartite synapse. 
If $n_i$ is the presynaptic neuron, $n_j$ is the postsynaptic neuron, and $a_i$ is an astrocyte, $s_t$ is defined as  

 \[ s_t := \{(n_i, n_j),(n_i,a_i),(a_i,n_j)\}. \]
\end{definition}  

Since not all directed synapses between neurons are necessarily associated with astrocytes, we define a type of configuration called one-one-directed configuration in the following.    

\begin{definition}
\label{def_one_one}
A one-one-directed configuration is represented by $s_d$ that is a set containing only one ordered pair. If $n_i$ is the presynaptic neuron and $n_j$ is the postsynaptic neuron, $s_d$ is defined as

\[ s_{d} := \{(n_{i},n_{j})\}. \] 

\end{definition}  

Gap junctions can also connect neurons and astrocytes \cite{fujii2017astrocyte,nagy2018electrical,semyanov2019spatiotemporal}. Gap junctions mainly allow the diffusion of ions and small molecules between neurons or between astrocytes \cite{fujii2017astrocyte,nagy2018electrical,semyanov2019spatiotemporal}. Since gap junctions bidirectionally connect neural-like cells, we represent their synaptic configurations as symmetric binary relations. We will call this synaptic configuration symmetric configuration instead of gap junction configuration to highlight that the information flow is bidirectional.

\begin{definition}
\label{def_gap_junction}
A symmetric configuration is represented by $s_s$ that is set containing two ordered pairs. If two neuron-like cells, $x_{i}$  and $x_{j}$, from the same cellular type (i.e. $x_{i}, x_{j}  \in  N \textrm{ or } x_{i}, x_{j} \in  A$) are connected by a gap junction, $s_s$ is defined as

 \[ s_{s} := \{(x_{i},x_{j}),(x_{j},x_{i})\}. \] 


\end{definition}

All these synaptic configuration are graphically explained in fig \ref{fig:1}.
\begin{note}
    In an ordered pair $(a,b)$, $a$ is the first coordinate and $b$ is the second coordinate.  
\end{note}
  
\subsection{$\Upsilon$-graph representation} 

Here, we define two representations of the collection of synapses of a neuron-glia network: $\Upsilon$-family and $\Upsilon$-multiset. But first, we define a set of synaptic configurations $S_X$ on the NL-set $X$ of a neuron-glia network. For the sake of clarity, the subscript of a set of synaptic configurations refers to the NL-set on which the synaptic configurations are defined. 
\begin{definition}
A set of synaptic configurations $S_X$ is a finite collection of sets of ordered pairs within $X \times X$, where $X$ is a NL-set, satisfying the following conditions: 

A1. All elements of $S_X$ are synaptic configurations $s_d$ (one-one-directed configuration), $s_s$ (symmetric configuration), or $s_t$ (tripartite configuration).

A2. There is at least one tripartite configuration $s_t$ in $S_x$. 

A3. All neuron-like cells of $X$ are a coordinate of an ordered pair that belong to at least one synaptic configuration of $S_X$.
\end{definition}

\begin{note}
A2 defines the minimal composition of a NL-set in the context of neuron-glia networks. This condition is necessary because there must be, at least, one tripartite synapses in a neuron-glia network. A3 guarantees that the elements of $X$ participate in at least one synaptic configuration of $S_X$. In other words, all neuron-like cells form at least one synapse.   
\end{note}

\begin{figure}[ht]
\centering
\includegraphics[scale=0.5] {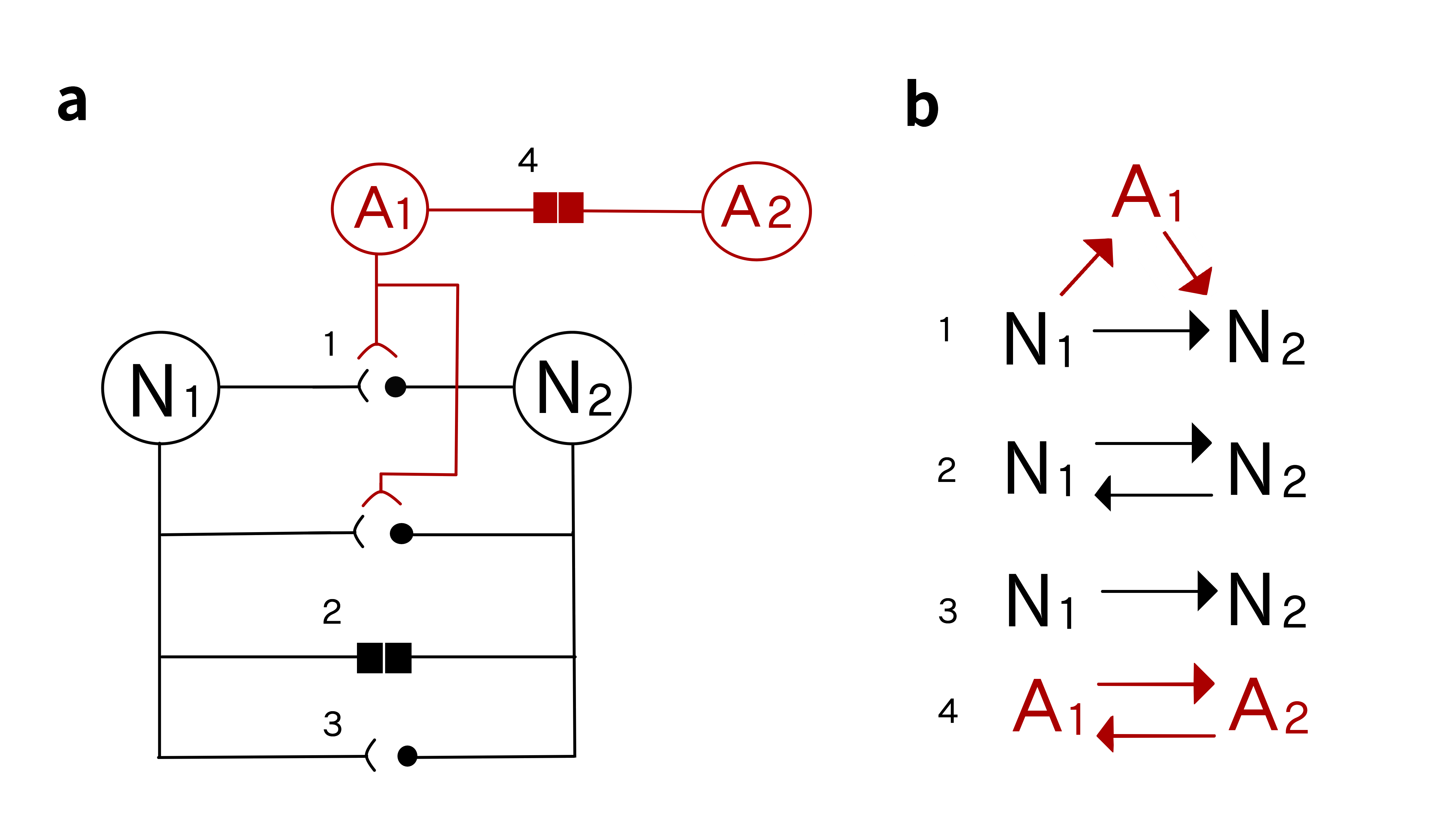}
\caption{a) Schematic representation of a neuron-glia network. $N_1$ and $N_2$ are two neurons, and $A_1$ and $A_2$ are two astrocytes. 1 is a tripartite synapse, 2 is a gap junction synapse between neurons, 3 is an one-one-directed synapse, and 4 is a gap junction between astrocytes. We show an additional non-labeled tripartite synapse to highlight that two single synapses can share the same synaptic configuration. b) Graphical representation of the directions of information flux inside synapses (synaptic configuration) (numbers refer to single synapses showed in a)) }
\label{fig:1}
\end{figure}

The ordered pairs within the elements of $S_X$ represent the directions of information flow inside a single synapse (see section \ref{shape} for a biological definition of single synapse or see fig \ref{fig:1}a for a diagram); however, in a network, single synapses can share a synaptic configuration. To represent this fact, we define the following concept. 
\begin{definition}

An $\Upsilon$-family $F$ is a finite indexed family $F: I \to S_X$  where $I$ is the set of single synapses of $X$ ($I \neq \varnothing$) such that $F(a)=b$ if and only if $b$ is the synaptic configuration of $a$. 
\end{definition}

$\Upsilon$-families relate single synapses with their synaptic configurations. Consequently, two single synapses $i$ and $j$ with synaptic configuration $s$ can be represented by the equations $F(i) = s$ and $F(j) =s$. In other words, $\Upsilon$-families are functions from real connections (single synapses) to mathematical objects (synaptic configurations). To facilitate the comparison of networks, we define the concept of $\Upsilon$-multiset which is a multiset represented by a count function that count how many single synapses share a synaptic configuration in an $\Upsilon$-family. 

\begin{definition}  \label{defupmulti}
An $\Upsilon$-multiset $A_x$ is a mset represented by a function $\text{Count}: S_X \to \mathbb{Z}^+ $, satisfying the following condition : $\forall U \in S_X$, $Count_{A_x}(U)= \textit{} \mid\! \{ i: F(i) = U \}\!\mid$ where $F$ is an $\Upsilon$-family $I \to S_X$. 
\end{definition}

 Since $\{ i: F(i) = U \}$ is the set of preimages under $F$ of $U$, its cardinality represents the number of occurrence of a synaptic configurations in the network. Now, given definition \ref{defupmulti}, we can define an $\Upsilon$-graph that represent a neuron-glia network as follows. 

\begin{definition}
 An $\Upsilon$-graph is a tuple $(X, A_x)$ where $X$ is a NL-set and $A_x$ is an $\Upsilon$-multiset represented by a function $\text{Count}: S_X \to \mathbb{Z}^+$. 
\end{definition}

A $\Upsilon$-graph $(X,A_x)$ is called \textit{connected} $\Upsilon$-graph if for every different pair of elements $x,y \in X$, there exists sequence $(a_n)_0^k$ where each pair of consecutive elements is part of a some relation (set of ordered pairs) $r \in A_x$ and $a_0=x \land a_k=y$. 

\subsection{$\Upsilon$-graph isomorphism} 

Similarly to multidigraph homomorphism (definition \ref{multidig-homo}), we define homomorphism for $\Upsilon$-graphs in definition \ref{upgrap-homo}. But, first, for the sake of clarity, we introduce the following notation, being $f$ a bijective function between two NL-sets and $s$ a set of ordered pairs of elements of the domain of $f$, if $s=\{(a,b)\}$, $s=\{(a,b),(a,b)\}$, or $s=\{(a,b),(b,c),(a,c)\}$, then, $s_f$ is $\{(f(a),f(b))\}$, $\{(f(a),f(b)),(f(b),f(a))\}$, or $\{(f(a),f(b)),$ $(f(b),f(c)),$ $(f(a),f(c))\}$, in that order. It is clear that $s$ and $s_f$ can be associated with an injective function; however, such a function will be omitted for simplicity.

\begin{definition}
\label{upgrap-homo}
Given two $\Upsilon$-graphs $\alpha = (X, A_x)$ and $\beta = (Y, A_y)$, the bijective function $f: X \to Y$ is an homomorphism if, for any element $s$ of $A_x^*$, the following condition is true.
\[ \text{Count}_{A_x}(s) = \text{Count}_{A_y}(s_f) \]

\end{definition}
As before, if $\alpha$ is homomorphic to $\beta$ and $\beta$ is homomorphic to $\alpha$, $\alpha$ and $\beta$ are isomorphic ($ \alpha \cong \beta$). 

In order to add functional relevance to the isomorphism concept, we create the term $\alpha$-isomorphic (represented as $\cong_\alpha$) which is defined as follows. 
\begin{definition}
The $\Upsilon$-graphs $\alpha = (X,A_x)$ and $\beta = (Y,A_y)$ are $\alpha-$isomorphic if the homomorphisms from $\alpha$ to $\beta$ and from $\beta$ to $\alpha$ only relate neurons to neurons and astrocytes to astrocytes. 
\end{definition}

This concept will also be applied to multidigraph isomorphisms if the vertex sets contains neurons and astrocytes. So, if  $x \cong_\alpha y$, then the homomorphisms from $x$ to $y$ and from $y$ to $x$ are bijective functions that relate neurons to neurons and astrocytes to astrocytes. 

\subsection{Unnesting an $\Upsilon$-graph}

Every $\Upsilon$-multiset is based on a NL-set (definition \ref{defupmulti}) and, for every $\Upsilon$-multiset, there is only one associated NL-set (see appendix \ref{secA1}, proposition \ref{proponeNLperomulti}). Clearly, this implies that for a given $\Upsilon$-multiset $A_x$, there exists one and only one associated connected $\Upsilon$-graph $(X,A_x)$. Now, let $pre-\phi$ be a function that relates $\Upsilon$-multisets with multisets of all the ordered pairs of all elements of the $\Upsilon$-multiset (see appendix \ref{secA2} for a detailed definition). Similarly to proposition proposition \ref{proponeNLperomulti}, we show that there exists only one NL-set per $pre-\phi(A_x)$ for any $A_x$ in proposition \ref{proponeNLperprephi} (see appendix \ref{secA3}). Therefore, for a given multiset  $pre-\phi(A_x)$, there exists one and only one associated multidigraph $(X,pre-\phi(A_x))$. Using $pre-\phi$, it is possible to construct another function called $\phi$ that relates $\Upsilon$-graphs to multidigraphs as it is shown in the following definition. 
\begin{definition}
\label{phi-def}
Being $\mathbf{P}$ a non-empty set of $\Upsilon$-graphs and $\mathbf{Q}$ a large enough non-empty set of multidigraphs, $\phi$ is a function defined as
\begin{equation}
\begin{array}{cccc}
\phi: &\mathbf{P}&\to&\mathbf{Q}\\
&x&\mapsto&y=\phi(x)
\end{array}
\end{equation}
if $x:=(X,A_x)$ and $y:=(V(y),E(y))$, then
$y$ is the image of $x$, i.e., $y=\phi(x)$, if and only if $X=V(y)$ and $E(y)=pre-\phi(A_x)$. 
\end{definition}

\begin{figure}
    \centering
    \includegraphics[width=0.6\linewidth]{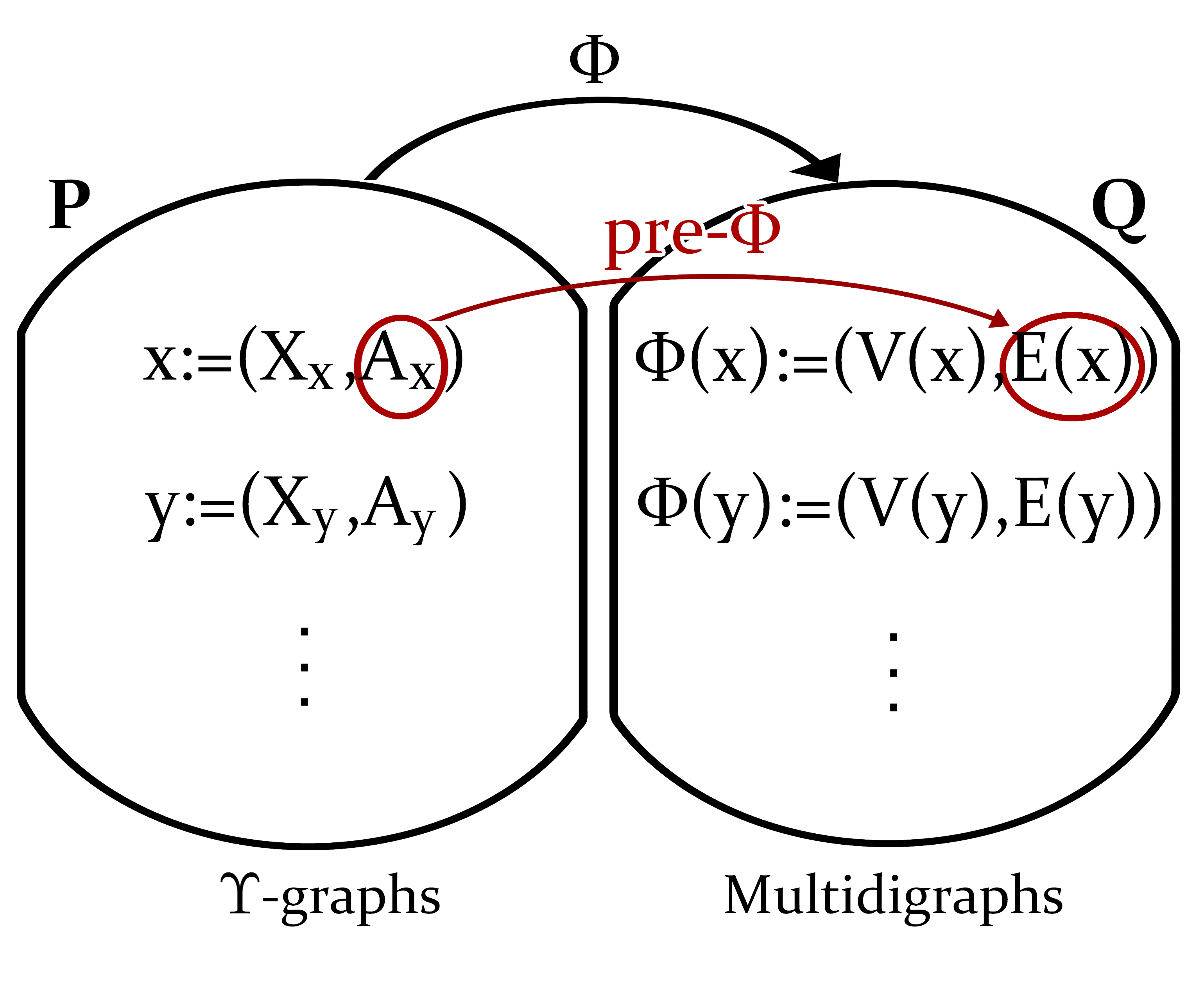}
    \caption{Venn diagram of a set of $\Upsilon$-graphs ($\mathbf{P}$) and a set of their associated multidigraphs ($\mathbf{Q}$), as well as the function $\phi$ that connects them. $\mathbf{P}=(x,y,...)$ and $\mathbf{Q}=(\phi(x),\phi(y),...)$} 
    \label{fig:2}
\end{figure}

This new function would represent the “unnested relation”, being the image the unnested form of the argument. Fig. \ref{fig:2} graphically explains the domain and range of $\phi$.

\subsection{Connecting $\Upsilon$-graphs and their unnested forms through matrices}

Let $x:=(X,A_x)$ be any $\Upsilon$-graph and $\phi(x):=(X,pre-\phi(A_x))$, its unnested form. The following proposition \ref{prop_sum} establishes a connection between $x$ and $\phi(x)$ in terms of repetition of synaptic configurations and ordered pairs.

\begin{proposition}
\label{prop_sum}
Let $\alpha\in pre-\phi(A_x)$ be any vertex of $\phi(x)$ (i.e., an ordered pair), if $s_1,s_2,...,s_k$ are all synaptic configurations (elements of $A_x$) that include $\alpha$ ($\alpha\in s_i$ for all $i\in[1,k]\cap \mathbb{Z}$), then
\begin{equation}
Count_{pre-\phi(A_x)}(\alpha)=Count_{A_x}(s_1)+ Count_{A_x}(s_2)+...+Count_{A_x}(s_k)
\label{eq-basic}
\end{equation} 
\end{proposition}
 Proposition \ref{prop_sum} states that, for any ordered pairs of the unnested form (multidigraph), the number of repetitions of the ordered pair is the summation of the number of its occurrences across all synaptic configurations in which it is present. This is formally proven in appendix \ref{proof_1}.   

If we assemble all the equations of the form ($\ref{eq-basic}$) of all the elements of $pre-\phi(A_x)^{*}$, then we form a system of equations that works as a more intuitive bridge between $x$ and $\phi(x)$ than $\phi$ itself. If $pre-\phi(A_x)^{*}$ has $m$ elements, the system of equations takes the following form. 
\begin{equation}
\label{eq-system}
    \begin{cases}
    Count_{pre-\phi(A_x)}(\alpha_1)=Count_{A_x}(a_1)+...+Count_{A_x}(a_k)\\
    Count_{pre-\phi(A_x)}(\alpha_2)=Count_{A_x}(b_1)+ ...+Count_{A_x}(b_{k^{'}})\\
    ...\\
    Count_{pre-\phi(A_x)}(\alpha_m)=Count_{A_x}(z_1)+...+Count_{A_x}(z_{k^{''}})\\
    \end{cases}
\end{equation}
where $\alpha_1\in a_1$, ..., $\alpha_1\in a_k$, $\alpha_2\in b_1$, and so on. Take not that $a_1$ is not necessarily different from $b_1$. We can go a step further to order the system of equations \ref{eq-system} and hence facilitate the operations.  In fact, if $A_x^{*}$ has $n$ elements ($s_1, s_2,...,s_n$), then the system of equations is equivalent to
\begin{equation}
\label{systemeq}
    \begin{cases}
    Count_{pre-\phi(A_x)}(\alpha_1)=c_{11}Count_{A_x}(s_1)+...+c_{1n}Count_{A_x}(s_n)\\
    ...\\
    Count_{pre-\phi(A_x)}(\alpha_m)=c_{m1}Count_{A_x}(s_1)+...+c_{mn}Count_{A_x}(s_n)\\
    \end{cases}
\end{equation}
Where each $c_{ij}$ represents the presence ($c_{ij}=1$) or absence ($c_{ij}=0$) of the $i$-th ordered pair ($\alpha_i$) in the synaptic configuration $s_j$ . It is possible to represent this later system of equations \ref{systemeq} with the following matrix equation:

\begin{equation} \label{matrix-eq}
    P=Ax
\end{equation}
which is amenable for any mathematical treatment. This matrix equation is called the $\Upsilon$-graph-multidigraph equation or $\Upsilon$-gm equation.  

The vector $P$ is a $m$ by 1 matrix. Each element represents the number of occurrences of an ordered pair in $pre-\phi(A_x)$ ($Count_{pre-\phi(A_x)}$).
The coefficient matrix $A$ is a $m$ by $n$ matrix with 1 and 0 as elements. 1 at the position $(i,j)$ implies that the ordered pair at the position $(i,1)$ of $P$ is included in the synaptic configuration $(j,1)$ of $x$. We can think in every column of the matrix as a synaptic configuration. As a consequence of definitions of the synaptic configurations (definition \ref{def_tri}-\ref{def_gap_junction}), a column representing a tripartite synapse has 1 at three positions, a column representing a gap junction has 1 at two positions, and a column representing a one-one-directed synapses has 1 at only one position.    
Finally, the vector $x$ is a $n$ by 1 matrix; each element represents the number of occurrences of a synaptic configuration in $A_x$ ($Count_{A_x}$). Note that all the elements of the matrices $P$, $A$, and $x$ are always non-negative integers.

There is an inherent challenge in the interpretation of $\Upsilon$-gm equations due to lack of a natural ordering of the synaptic configurations and their elements (information flows), i.e., any ordered pair could be labeled as $\alpha_1$ or $\alpha_n$, and any synaptic configuration could be labeled as $s_1$ or $s_n$. Hence, the same network could be represented in a finite number of ways which is the factorial of the number of synaptic configurations times the factorial of number of information flows. Therefore, when we compare two different networks based on their $\Upsilon$-gm equations, some counterintuitive situations may arise. We explore these situations in the following.

\begin{enumerate}
\item \textbf{False negative}: Given two $\Upsilon$-gm equations $P=Ax$ and $Q=By$ that represent two neuron-glia networks $\alpha$ and $\beta$. $\alpha=\beta$ does not necessarily imply that $P=Q$, $A=B$, or $x=y$. In other words, \textbf{different} $\Upsilon$-gm equations can represent the same neuron-glia network.
\item \textbf{False positive}: Given two $\Upsilon$-gm equations $P=Ax$ and $Q=By$ that represent two neuron-glia networks $\alpha$ and $\beta$. $P=Q$ does not guarantee that $A=B$ or $x=y$. 
\item \textbf{True comparison}: Given two $\Upsilon$-gm equations $P=Ax$ and $Q=By$ that represent two neuron-glia networks $\alpha$ and $\beta$. If $A=B$ and $x=y$, then $\alpha$ and $\beta$ are isomorphic. (see the last part of the \textit{proof} of proposition \ref{secondmainprop})
\end{enumerate}

\section{Multidigraphs can misrepresent neuron-glia networks}

In this section, we will prove that the
existence of isomorphism between the unnested
forms of $\Upsilon$-graphs does not imply that the
networks they represent share the same pattern. To do so, we found pairs of non-isomorphic $\Upsilon$-graphs with isomorphic unnested forms. In the following, we will show four scenarios where isomorphism of the unnested form does not implies isomorphism of the $\Upsilon$-graphs. Keep in mind that $\mathbf{P}$ is a non-empty set of $\Upsilon$-graphs. At the end of the section (proposition \ref{secondmainprop}), we will prove, in the general case, that the isomorphism between the unnested
forms of $\Upsilon$-graphs implies the isomorphism of  their $\Upsilon$-graphs under some specific conditions. 


\begin{enumerate}
    \item All types of synaptic configuration can be present (proposition \ref{prop_case1}). 
    \item There are only tripartite configurations (proposition \ref{prop_case2}).
    \item There are only tripartite configurations and one-one-directed configurations (proposition \ref{prop_case3}).
    \item There are only tripartite configurations, and symmetric configurations (gap junctions) (proposition \ref{prop_case4}).
    
\end{enumerate}

To prove propositions \ref{prop_case1}-\ref{prop_case4}, we use theoretical examples where the isomorphism between the unnested forms does not imply the isomorphism between $\Upsilon$-graphs. For example, to prove proposition \ref{prop_case1}, we present two small network motifs (represented by matrices $A_{5 \times 2}$ and $B_{5 \times 3}$) formed by two neurons and one astrocyte linked by one tripartite synapse (note that matrix $A$ and $B$ have 1 as entry of first three rows of the first column, representing a tripartite synapse) together with two one-one-directed synapses (two columns with 1 at one position in $A$) or one gap junction(1 column with 1 at two positions in $B$). We prove that the unnested form of of these different motifs is isomorphic, meaning that it is possible to find non-isomorphic neuron-glia networks with isomorphic unnested forms. We provided a formal explanation in the in the following proof. In this proof, we establish basic mathematical notation for the following proofs because it is easier to explain them in a simple example rather than in more complicated examples or in the general case (proposition \ref{secondmainprop}). Keep in mind that all examples showed bellow are theoretical examples and when we say “there are”, it means “it is theoretical possible” since the examples do not come from a database. 

\begin{proposition}
\label{prop_case1}
There are two $\Upsilon$-graphs $a$ and $b$ (all synaptic configuration allowed) such that the unnested form of $a$ and $b$ are $\alpha$-isomorphic ($\phi(a)\cong_\alpha\phi(b)$) but $a$ and $b$ are non-isomorphic ($\neg (a\cong b)$).
\end{proposition}
\begin{proof}
Let's define the elements of the proposition as follows, $a:=(X, A_x)$, $b:=(Y,A_y)$,
$\phi(a):=(X,pre-\phi(A_x))$, and $\phi(b):=(Y, pre-\phi(A_y))$.
Now, since $ a\cong b$ if and only if $a$ is homomorphic to $b$ and $b$ is homomorphic to $a$, then
we define its negation as
\begin{equation}
\label{neg-cong}
\neg(a\cong b) \iff a\textit{ is non-homomorphic to }b \lor b\textit{ is non-homomorphic to }a
\end{equation}

Since $\phi(a)\cong_{\alpha}\phi(b)$ implies that there exists $\Upsilon$-gm equations $P=Ax$
and $Q=B_{0}y$ such that they represent the neuron-glia networks of $a$ and $b$, respectively,
and $P = TQ$ (lemma \ref{lemma_1}, appendix \ref{lemma_case1}), we see that $TQ = TB_{0}y$ represents the second neuron-glia network. Being $B = TB_{0}$, we will use $P=Ax$
and $P=By$ to refer to the $\Upsilon$-gm equations
of the neural networks represented by $a$ and $b$, respectively.
Then, using the language of $\Upsilon$-gm equations, in some cases $A\neq B$ or $x\neq y$ will imply
(\ref{neg-cong}) while holding $\phi(a)\cong_{\alpha}\phi(b)$ true. This is the approach that we will follow to prove the first proposition.

Let A and B be the following matrices,
        \begin{center}
    $A=$
    $\begin{bmatrix}
    1&0\\
    1&0\\
    1&0\\
    0&1\\
    0&1\\
    \end{bmatrix}
    $
     and 
    $B=$
    $\begin{bmatrix}
    1&0&0 \\
    1&0&0 \\
    1&0&0 \\
    0&1&0 \\
    0&0&1 \\
    \end{bmatrix}
    $
    \end{center}
    If $x=[\alpha,\beta]^T$ and $y=[\alpha,\beta,\beta]^T$, then $Ax$ and $By$ are equal to $P=[\alpha,\alpha,\alpha,\beta,\beta]^T$.
We realize that it is impossible to establish $a\cong b$ as $A_x^{*}$ and $A_y^{*}$ have
different cardinality. Therefore, we have proved a case in which $\phi(a)\cong_{\alpha}\phi(b)$
and $\neg(a\cong b)$. This case can be easily generalized to networks containing $a$ and $b$
as sub-networks.
    
\end{proof}
The previous instance ($C_1$) reflects the lack of “resolution power" of models that only use ordered pairs as elements such 
as digraphs or multidigraphs. Indeed, at a physiological, the neuron-glia networks represented by $a$ and $b$ will not show the same behaviour
since two one-one-directed synapses are not equivalent to a gap junction. However, due to the relative difficulty to find gap junctions in networks
due to technical limitations, we present the following instance of $C$ that does not posses them. To prove 
 proposition \ref{prop_case2}, we find that two solutions for an $\Upsilon$-gm equation (defined in \ref{matrix-eq}), proving that there can be  non-isomorphic $\Upsilon$-graphs with isomorphic unnested forms even when gap junctions are not allowed. We follow the same approach to prove proposition \ref{prop_case2} - \ref{prop_case4}. 

\begin{proposition}
\label{prop_case2}
There are two $\Upsilon$-graphs $a$ and $b$ (only tripartite configurations allowed) such that the unnested form of $a$ and $b$ are $\alpha$-isomorphic ($\phi(a)\cong_\alpha\phi(b)$) but $a$ and $b$ are non-isomorphic ($\neg (a\cong b)$).
\end{proposition}
\begin{proof}
    Using the same notation as in the previous proof, from this case on, we will use a basic concept of linear algebra to prove the existence of $\Upsilon$-graphs
that hold the statements.
We can make the coefficient matrices equal ($A=B$), and then find different vectors that
solve $P=Ax$. Each solution (vector) corresponds to a different $\Upsilon$-graph; therefore, after proving that
the system has multiple solutions we will show that they represent non-isomorphic $\Upsilon$-graphs. In this
way, we start showing that the null space of $A$ has more solutions than the zero vector. Let $A$ be the
following matrix,
    \begin{center}
    $A=$
    $\begin{bmatrix}
    1&1&0&0&0&0&0&0 \\
    1&0&1&0&0&0&0&0 \\
    1&0&0&1&0&0&0&0 \\
    0&1&0&0&1&0&0&0 \\
    0&1&0&0&0&1&0&0 \\
    0&0&1&0&0&1&0&0 \\
    0&0&1&0&0&0&1&0 \\
    0&0&0&1&1&0&0&0 \\
    0&0&0&1&0&0&1&0 \\
    0&0&0&0&1&0&0&1 \\
    0&0&0&0&0&1&0&1 \\
    0&0&0&0&0&0&1&1 \\
    \end{bmatrix}$
    \end{center}
$Ax=0$ has non-zero solutions such as $s=[1,-1,-1,-1,1,1,1,-1]^T$ and any vector $ks$ where $k$ is a positive
integer. This implies that $P=Ax$ has either 0 or infinite solutions. 

Now we will take a case in point, $x_1=[3,1,1,1,3,3,3,1]^T$ and $x_2=[2,2,2,2,2,2,2,2]^T$, we see that
$Ax_1$ and $Ax_2$ are equal to the vector $P=[4,4,4,4,4,4,4,4]^T$. It is clear that the $\Upsilon$-graphs 
that are represented by $x_1$ and $x_2$ are not isomorphic since it is impossible to establish an
homomorphism between $a$ and $b$ since $3\neq 2$ and $1\neq 2$.\\
In ordinary words, this means that are neuron-glia networks that share the same $Count$ of information flows (elements of the synaptic configurations) while holding different $Count$s of synaptic configurations. Finally, networks containing $A$ as a part of the coefficient matrix may also satisfy the general statement 
$\phi(a)\cong_{\alpha}\phi(b)\land\neg(a\cong b)$.

\end{proof}
Since the previous matrix equation may be a little hard to find in a real network due to combinatory of arrangements
between seven tripartite synapses, we present the following motif that is most likely to be part of a neuron-glia network because it has fewer tripartite and some one-one-directed synapses.
\begin{proposition}
\label{prop_case3}
There are two $\Upsilon$-graphs $a$ and $b$ (symmetric configuration not allowed) such that the unnested form of $a$ and $b$ are $\alpha$-isomorphic ($\phi(a)\cong_\alpha\phi(b)$) but $a$ and $b$ are non-isomorphic ($\neg (a\cong b)$).
\end{proposition}
\begin{proof}
Continuing with the same notation, likewise the above case, let $A$ be the following matrix,
    \begin{center}
    $A=$
    $\begin{bmatrix}
    1&1&0&0&0&0&0&0\\
    1&0&0&0&1&0&0&0\\
    1&0&1&0&0&0&0&0\\
    0&0&1&0&0&0&1&0\\
    0&0&0&0&1&0&1&0\\
    0&0&0&0&0&0&1&1\\
    0&0&0&0&1&1&0&0\\
    0&0&1&1&0&0&0&0\\
    \end{bmatrix}
    $
    \end{center}
    Here, the solutions of $Ax=0$ are of the form $ks$, where $s=[$1, -1, -1, 1, -1, 1, 1, -1$]^{T}$ and $k$ is a positive integer. As an example, let $x_1=[3,1,1,3,1,3,3,1]^T$ and $x_2=[2,2,2,2,2,2,2,2]^T$, $Ax_1$ and $Ax_2$ are equal to $P=[4,4,4,4,4,4,4,4]^T$. Using the same reasoning as above, the $\Upsilon$-graphs that are represented
by $x_1$ and $x_2$ are not isomorphic.
\end{proof}
The following case disregards one-one-directed synapses altogether to show that even in such particular scenario 
our proposition is still true.
\begin{proposition}
\label{prop_case4}
There are two $\Upsilon$-graphs $a$ and $b$ (one-one-directed configuration not allowed) such that the unnested form of $a$ and $b$ are $\alpha$-isomorphic ($\phi(a)\cong_\alpha\phi(b)$) but $a$ and $b$ are non-isomorphic ($\neg (a\cong b)$).
\end{proposition}
\begin{proof}
    Following the same notation, in this scenario, the coefficient matrix of interest $A$ is showed in appendix \ref{secA4}. The solutions of $Ax=0$ are of the form $ks$, where $s= [-1, 1, 1, -1, -1, 1, 1, -1, -1, 1, 1, -1, -1, -1, 1, 1, 1, 1, -1, -1, 1, -1]^{T}$ as a solution for $Ax=0$. and $k$ is a positive integer.
As an example, let $x_1=[1,3,3,1,1,3,3,1,1,3,3,1,1,1,3,3,3,3,1,1,3,1]^T$ and $x_2=[2,2,2,2,2,2,2,2,2,2,2,2,2,2,2,2,2,2,
2,2,2,2]^T$, $Ax_1$ and $Ax_2$ are equal to $P=[4,4,4,4,4,4,4,4,4,4,4,4,4,4,4,4,4,4,4,4,4]^T$. Once again, the $\Upsilon$-graphs that are represented by $x_1$ and $x_2$ are not isomorphic. 
\end{proof}

In propositions \ref{prop_case1}-\ref{prop_case4}, we realized that for a pair of representations of a neuron-glia network $a$ and $\phi(a)$,
where $a$ is a $\Upsilon$-graph and $\phi(a)$, its unnested form, it is safer to work with $a$ rather than
$\phi(a)$ since it is theoretically possible to lose information during the unnesting of the $\Upsilon$-graph.
It is important to mention that some of the showed examples may not exist in real networks but they prove that it is not safe to assume that the isomorphism between the unnested forms ($\phi(a)\cong_\alpha\phi(b)$) implies the isomorphism between $\Upsilon$-graphs ($a\cong b$).
However, we conjectured that 
\begin{equation}
    \label{iso_statment}
\phi(a)\cong_\alpha\phi(b)\implies a\cong b
\end{equation}

holds under some specific conditions. The conditions in which \ref{iso_statment} is true are defined in the following paragraph and, in proposition 
\ref{secondmainprop}, we prove that these conditions are sufficient for \ref{iso_statment} to be true. 

$Q_1$ and $Q_2$ are conditions applied over two $\Upsilon$-graphs $a$ (with $A_a$ as its $\Upsilon$-multiset) and $b$ (with $A_b$ as its $\Upsilon$-multiset). Being $Ax=P$ and $B_0y=Q$, the $\Upsilon$-gm equations related to $a$ and $b$, respectively, if $rank(A)$ and $rank(B_0)$ are the ranks of $A$ and $B_0$, respectively, then
$Q_1\equiv rank(A)=\mid A_a^{*}\mid \land \ rank(B_0)=\mid A_b^{*}\mid$.
This implies that all the columns of $A$ and $B_0$ are linearly independent in their matrices. Being $T$ a row exchange matrix, $Q_2$ means that if $x =Ty$, then $A = TB_0$.
\begin{proposition}
For every pair of $\Upsilon$-graphs, $a$ and $b$, such that they hold $Q_1$ and $Q_2$ and the unnested forms of $a$ and $b$ ($\phi(a)$ and $\phi(b)$) are $\alpha$-isomorphic, then $a$ and $b$ are isomorphic. 

\label{secondmainprop}
\end{proposition}


\begin{proof}
$Q_1$ guarantees that both $\Upsilon$-gm equations have either 1 or no solutions since the null space of the coefficient matrices $A$ and $B_0$ are trivial, i.e., only contain the zero vector as a solution. Now, given the definition of $\phi(a)$ (definition \ref{phi-def}), the equations are always solvable, therefore they have 1 and only 1 solution.\\
Since Lemma \ref{lemma_1}, $P=TQ$ where $T$ is a row exchange matrix. Therefore, we multiply $T$ to both sides of the $\Upsilon$-gm equation of $b$. The result, $TB_0y=P$, is also a valid $\Upsilon$-gm equation that represents the relationship between $b$ and $\phi(b)$. Let $TB_0=B_1$, we have the following equations

$$
\begin{cases}
Ax=P \\
B_1y=P
\end{cases}
$$
Now, due to condition $Q_2$, $B_1=A$; therefore, $Ax=Ay$. 
Now, realize that since $A$ has a trivial null space, $Ax=Ay$ implies that $x=y$. This is explained with the following reasoning,

\begin{equation*}
\begin{array}{ccl}
Ax=Ay & \implies & A(x+(-1)y)=Ax+A(-y)\\
& \implies & A(x-y)=Ay+A(-y)=Ay-Ay\\
& \implies & A(x-y)=0\\
& \implies & x-y=0\\
& \implies & x=y
\end{array}
\end{equation*}
The fourth step is based on the fact that $A\alpha=0$ has only one solution which is $\alpha=0$. To summarize, $Ax=P$ is the $\Upsilon$-gm equation of the $\Upsilon$-graphs $a$ and $b$. The last step is to show that this implies that $a\cong b$.
Let's define both $\Upsilon$-graphs as $a:=(X,A_x)$ and $b:=(Y,A_y)$. Keep in mind that $a$ and $b$ share the same types ($A=A$), number  ($A=A$), and repetition of synapses ($x=x$). Now, let's choose a function among all the bijections that can be defined between $X$ and $Y$ such that
$$\forall a,b\in X, (a,b)\in pre-\phi(A_x)\implies (f(a),f(b))\in pre-\phi (A_y) $$
We can verify that $\alpha:=\{(a,b):a,b\in X\}\in A_x \implies \alpha_f:=\{(f(a),f(b)):f(a),f(b)\in Y\}\in A_y$ since every ordered pair from $pre-\phi(A_x)$ is 1-1 associated with a pair of $pre-\phi(A_y)$. However, since $a$ and $b$ share the same coefficient matrix ($A$) and $Count$ vector ($x$), the following is true for every synapse of $A_x$, $Count_{A_x}(\alpha)=Count_{A_y}(\alpha_f)$, in other words, $f$ is an homomorphism from $X$ to $Y$. We can analogously define an homomorphism from $Y$ to $X$, concluding that $a\cong b$.

\end{proof}

\section{Modeling 
 the network shape}
\label{shape}
Nowadays, a huge amount of information about neural shape and network connectivity is available in open databases \cite{clements2020neuprint, vogelstein2018community, xu2013computer}. However, the mathematical tools to study neuronal shape are still under development \cite{kanari2018topological, li2017metrics}. Natural structure theory could contribute to this research field by providing a theoretical framework to study the shape of neurons, astrocytes, or even networks \cite{cabrera2004natural2,cabrera2004natural1}. In this section, we use a mathematical formalism (definition \ref{NEdef}) that represents the concept of a natural structure \cite{cabrera2004natural2,cabrera2004natural1} to model the morphology of a neuron-glia network.   

A connected neuron-glia network is evidently a natural structure, where its elements are neural-like cells and single synapses are NE-connections \cite{cabrera2004natural2,cabrera2004natural1}. In general, a single synapse is composed of a presynaptic membrane, an astrocyte membrane (if it is tripartite), a postsynaptic membrane, and a perisynaptic matrix (if it is chemical). The synaptic cleft is a region of the space filled of perisynaptic matrix which is composed of proteoglycans and proteins \cite{pinter2020chondroitin,jang2017synaptic, rudenko2019neurexins}. This matrix mediates trans-synaptic adhesion (NE-connection role) and regulates the synaptic transmission (communication role) \cite{jang2017synaptic, wierda2020soluble}. Thus, a single synapse can be thought of as an structure. In this section, we roughly model a neuron-glia network, applying part of the mathematical definition of natural structures (definition \ref{NEdef}) from Cabrera-Febola, 2023-in preparation for publication. 

\begin{definition}
\label{NEdef}
(Cabrera-Febola, 2023-in preparation for publication) A natural structure (NE) is a triple $(X,D,C)$, where $X$ is a collection of the elements of the NE ($\varnothing \notin X$), $C$ is a family of collections of NE-connections, and $D$ is a distribution of the elements, being a pair ($S$,$\{ m_s \hookrightarrow S : m \in X, s = 1,2,...,v, v \in \mathbb{N} \}$), where $S$ is the “known" natural space and $ m_s \hookrightarrow S$ stands for $m_s$ exists along $S$. 
\end{definition}

In our case, the natural structure model of a connected neuron-glia network (connected $\Upsilon$-graph) whose synaptic collection is represented by a $\Upsilon$-family $F: I \mapsto S_X$ is $(X,\mathbb{D},I)$ where $X$ is a set of neuron-like cells (astrocytes and neurons) and $I$ is a set of single synapses (NE-connections). In order to define $\mathbb{D}$ which models the distribution of elements, we define a distribution function $g: X \cup I \to P $ where $P$ is a family of subsets of $\mathbb{R}^3$ such that any element $x$ of the domain of $g$ is distributed in all the points of $g(x)$ and nowhere else. Hence, we can define $\mathbb{D}$ as $(\mathbb{R}^3,g)$ to model the spatial distribution of the network. 

Neuron-like cells have not cavities, so we can assume that $g(x)$ with $x \in X$ is homeomorphic to a solid ball. On the other hand, $g(s)$ with $s \in C$ is the space where neurotransmission occurs, thus it can be assumed to be also homeomorphic to a solid ball. Additionally, we can state that synapses are partially composed of membranes from all implied neuron-like cells, so $g(s) \cap g (x_i) \cap g(x_j) \cap ... \cap g(x_z)\neq \varnothing$ if and only if $s \in I$ and $F(s)$ is a relation on $\{x_i,x_j,..., x_z\}$. Therefore, since the $\Upsilon$-graph is connected, it is possible to conjecture that the $\{\bigcup g(e): e \in X \cup I \}$ is a solid. There should be more topological and geometrical properties of neuron-glial networks; however, this topic is out of the scope of our article.

\section{Discussion}

Neurons has been the focus of two hundreds of years of neuroscience research \cite{swanson2016cajal}. However, experimental evidence suggests that astrocytes play an important role in neuronal dynamics, making it necessary to deepen our understanding of how astrocytes network influences neural phenomena \cite{chever2016astroglial, deemyad2018astrocytes, lee2014astrocytes}. Although there were some attempts to model neuron-glia network dynamics, they lack of a formal network representation \cite{chan2017role,tang2013information,wade2011bidirectional}. To fill this lack, in this study, we present a formal representations of a neuron-glia network, the $\Upsilon$-graph.

An $\Upsilon$-graph is defined as a tuple $(X,S_X)$ where $X$ is a set of neural-like cells and $S_X$ a multiset of synaptic configurations. In our framework, synaptic configurations are sets of ordered pairs that represent the information flux inside a synapse. Since neurons can establish many synapses with the same information flux, i.e. the same synaptic configuration, $S_X$ is a multiset whose counting function allows us to know the number of occurrence of a synaptic configurations in the network. 

Previous works on neural networks had not considered each synapse as an individual element but the collection of synapses as the edge of the digraph or graph \cite{reimann2017cliques,scheffer2020connectome,towlson2013rich}. However, graph-like representations can misrepresent a neuronal network. For example, if a neuronal network has electrical synapses, each electrical synapse is equivalent to two chemical synapses with inverse directions in a digraph. In contrast, our work presents a new mathematical object, $\Upsilon$-graph which consider each single synapses (electrical, chemical, or tripartite synapse) as an element of the network.

An alternative representation that would address the issue of the multiple types of synapses in the brain would be a multilayer network \cite{de2023more}. In this mathematical object, a layer can represent a type of synapses, solving the misrepresentation problem discussed in the previous paragraph. However, multilayer networks like digraphs assign arcs between two vertices, thus misrepresenting the triadic nature of tripartite synapses. This misrepresentation is an issue when two different networks are represented by the same mathematical objects. In proposition \ref{prop_case2}, we found a theoretical example of multidigraph that can represent multiple different neuron-glia networks with only tripartite synapses. This misrepresentation can not be fixed by using multilayer networks because there is only one type of synapse in our example.

In addition to $\Upsilon$-graphs, there are other mathematical objects that can represent the information flow in triadic relations: directed hypergraphs \cite{gallo1993directed} and abstract directed simplicial complex \cite{reimann2017cliques}. The problem with directed hypergraphs is that a vertex can not be part of the input and outputs of an hyperarc, meaning that, in our context, a neuron-like cell (i.e. astrocyte or neuron) could only send information but not receive information throughout the same synapse. This is not the case for tripartite synapses because the astrocyte receives information from the presynaptic neuron and sends information to the postsynaptic neurons \cite{gonzalez2019gliotransmission, scimemi2019role, verkhratsky2020nmda, jennings2017dopamine}. The case of simplicial complexes need further study to prove wether it is always possible to built a simplicial complex from a $\Upsilon$-graph or not. Simplicial complexes are used to study the algebraic topology of networks \cite{greening2015higher,reimann2017cliques, sizemore2018cliques, sizemore2019importance} while graph-like objects are used to find network motifs and to make graph measures such as between centrality and modularity \cite{bassett2017network}.  

Since graph representations are widely used in neuroscience, we defined a function ($pre-\phi$, see appendix \ref{secA2}) that unnests $\Upsilon$-graphs into multidigraphs. However, using linear algebra, we proved that the isomorphism between unnested forms does not guarantee isomorphism between their respective $\Upsilon$-graphs in several scenarios (propositions \ref{prop_case1}-\ref{prop_case4}).
Therefore, multidigraph representations can misrepresent a neuron-glia networks. Interestingly, we also found that if the coefficient matrices and the vectors of synaptic configurations of the two $\Upsilon$-graphs are equal after interchanging rows, and the coefficient equations' columns are linearly independent, the isomorphism between unnested forms guarantees the isomorphism between $\Upsilon$-graphs. Under these specific conditions, we can use graph theoretical tools developed for multidigraphs without misrepresentation issues. 

In this work, we also sketched a novel approach to model the network shape. As $\Upsilon$-family represents each single synapse as a distinguishable element, we use this object to create a natural structure model. This natural structure model is a tuple $(X,\mathbb{D},I)$ where $X$ is a set of neural-like cells, $I$ is a set of single synapses,and $\mathbb{D}$ is a tuple $(\mathbb{R}^3,g)$ where $g$ is the distribution function. In our model, single synapses and neuron-like cells are homeomorphic to solid balls. We conjecture that the portion of space where neuron-like cells and synapses are distributed is also a solid. This hypothetical solid could be studied applying differential geometry or algebraic topology which open new ways for studying structural connectomes. 

In conclusion, we have developed a mathematical framework for working with connectivity structure of neuron-glia networks ($\Upsilon$-graphs, conectedness, isomorphism, and other constructs.) We have also connected the concept of $\Upsilon$-graphs with graph-like objects, multidigraphs, and studied their relations ($\phi$ function and matrix equation). On the other hand, we have sketched a natural structure model to represent the spatial structure of neuron-glia networks. 

\section{Acknowledgments}

We thank D. A. Pacheco (Harvard University) and Leonardo Torres (Max Planck Institute for Mathematics in the Sciences) for their helpful comments and partial manuscript proofreading.

\section{Declarations}

\subsection{CRediT author statement}

Marco Peña-Garcia: Conceptualization, Methodology, Investigation, Writing - Original Draft. Francesco Peña-Garcia: Methodology, Investigation, Writing - Original Draft. Nelson Castro: Conceptualization, Methodology, Writing - Review \& Editing, Supervision. \hspace{10cm} Walter Cabrera-Febola: Writing - Review \& Editing, Supervision.

\subsection{Competing interests}
The authors have no competing interests that are relevant to the content of this article.

\noindent


\section{Supplementary section}

\subsection{One NL-set per $\Upsilon$-multiset}\label{secA1}

In this appendix, we prove that there is only one NL-set per $\Upsilon$-msets. First, lets define a relation $\diamond$ between $\Upsilon$-msets and NL-sets as follows
$$\alpha\diamond \beta \iff (\alpha^{*}\subset \mathcal{P}(\beta\times\beta)\land(\forall x\in\beta \exists y\in\alpha:x\in Dom(y)\lor x\in Ran(y))) $$
Now, let $\mathbf{M}$ be an arbitrary non-empty set of $\Upsilon$-msets and $\mathbf{L}$, an arbitrary non-empty set of NL-sets, then  the following is true.
\begin{proposition} \label{proponeNLperomulti}
$$ \forall A\in\mathbf{M}\forall X,Y\in\mathbf{L}, ((A\diamond X\land A\diamond Y)\implies X=Y) $$
\end{proposition}
\begin{proof}
We will proceed with a \textit{reductio ad absurdum} argument. Let $P$ be the proposition, then
$$\neg P\equiv \exists A\in\mathbf{M}\exists X,Y\in\mathbf{L},(A\diamond X\land A\diamond Y)\land X\neq Y $$
\begin{equation*}
\begin{array}{ccc}
\neg P & \implies & ((\exists x\in X:x\not\in Y)\lor (\exists y \in Y:y\not\in X) )\\
& \implies &
((\exists \alpha\in A^{*}:(x\in Dom(\alpha)\lor x\in Ran(\alpha))\land \alpha\not\in\mathcal{P}(Y\times Y))\lor\\
&&(\exists \beta\in A^{*}:(y\in Dom(\beta)\lor y\in Ran(\beta))\land \beta\not\in\mathcal{P}(X\times X)))\\
& \implies & \neg(A\diamond Y)\lor\neg(A\diamond X) \\
& \implies & \neg(A\diamond X\land A\diamond Y)\\
& \implies & \neg((A\diamond X\land A\diamond Y)\land X\neq Y)\\
\end{array}
\end{equation*}
\begin{flushright}
$(\Rightarrow\!\Leftarrow)$
\end{flushright}
\begin{center}
$\therefore \forall A\in\mathbf{M}\forall X,Y\in\mathbf{L}, ((A\diamond X\land A\diamond Y)\implies X=Y) $
\end{center}
\end{proof}
You can also understand this proof as ``there is no $A$, $X$, and $Y$ such that $\neg P$ is true, it follows that if $\mathbf{M}$ and $\mathbf{L}$ are non-empty, then they hold $P$". Remember that $\neg (\neg P)\equiv P$.

\subsection{Defining $pre-\phi$ function}
\label{secA2}

In this appendix, we formally define the $pre-\phi$ function in a constructive way as follows.

Let $\mathbf{A}$ be an arbitrarily large set of finite multisets and $\mathbf{A}^m$ an arbitrarily large set of sets. For any $A\in \mathbf{A}$ and $B\in \mathbf{A}^m$, we say $A\sim B$ if and only if the following holds.

\begin{enumerate}
    \item $\forall a\in B,a\in A^*\times \mathbb{N}$.
    \item $\forall b\in A^* \exists a\in B:\{b\}\in a$.
    \item $\forall a \in B,\{c\}\in a\implies\{c,\text{Count}_A(c)\}\in a$.
\end{enumerate}
Being $c$ any element, i.e., a free variable of the statement. The first condition means that $B$ is a set of ordered pairs $(b,n)$ where $b$ is an element of $A^{*}$ and $n$, a natural number. The second condition implies that every element of $A^{*}$ is represented in $B$. The final condition defines $n$ for every $(b,n)$ as the $Count$ of $b$ in $A$. Since every $b\in A^{*}$ has only one $Count$, every $b$ is associated with one and only on $(b,n)$ in $B$.

Since $\forall A,B\in \mathbf{A}, C\in \mathbf{A}^m,(A\sim B\land A\sim C\implies B=C)$, and $\mathbf{A}$ and $\mathbf{A}^m$ are sets, the relation $\sim$ defines a function from $\mathbf{A}$ to $\mathbf{A}^m$. We denote this function as $\pi$.
\begin{equation*}
    \begin{array}{rl}
        \pi:\mathbf{A}\to & \mathbf{A}^m\\
        A\mapsto & B=\pi(A)
    \end{array}
\end{equation*}
where $B=\pi(A)\iff A\sim B$. It is clear that $\pi$ is injective.\\
There exists a subset of $\pi(\mathbf{A})$ whose elements has the following form,
$$ B:=\{(\alpha,n):\alpha\in S_X\land n\in \mathbb{N}\} $$
where $S_X$ is a set of binary relations on subsets of some NL-set $X$. This subset of $\pi(\mathbf{A})$ will be denoted as $S^{\pi}\subset \pi(\mathbf{A})$.\\

Let $\mathbb{U}$ be an arbitrary large set of sets of sets, we define a function $\delta$ from $S^{\pi}$ to $\mathbb{U}$, i.e.,
\begin{equation*}
    \begin{array}{rl}
        \delta:S^{\pi}\to & \mathbb{U}\\
        B\mapsto & C=\delta(s^{\pi})
    \end{array}
\end{equation*}
$\delta$ is defined as
$$ C=\delta(B) \iff b_1 \land b_2 $$
where
\begin{enumerate}
\item[$b_1\equiv$] $\forall (\alpha,n)\in B \exists c\in C: (a\in \alpha\implies (a,n)\in c)$.
\item[$b_2\equiv$] $|B|=|C|$.
\end{enumerate}
Keep in mind that $B=\pi(A)$ for some $A\in \mathbf{A}$, i.e., $C=\delta(\pi(A))$.\\
For the next step, we will define a sum operator $+^{c}$ as
\begin{equation}
\label{plus_c_def}
    \begin{array}{rl}
        +^{c}:\pi(\mathbf{A})\times\pi(\mathbf{A})\to &\pi(\mathbf{A}) \\
        (c_1,c_2)\mapsto & c_3=c_1+^{c}c_2
    \end{array}
\end{equation}
This operation is defined as
$$ c_3=c_1+^{c}c_2 \iff d_1 \land d_2\land d_3\land d_4 $$
where
\begin{enumerate}
\item[$d_1$] $(a,n)\in c_1\land (a,m)\in c_2\implies (a,m+n)\in c_3$.
\item[$d_2$] $\forall m\in\mathbb{N},(a,n)\in c_1\land (a,m)\not\in c_2 \implies (a,n)\in c_3$.
\item[$d_3$] $\forall m\in\mathbb{N},(a,n)\in c_2\land (a,m)\not\in c_1 \implies (a,n)\in c_3$.
\item[$d_4$] If $m_1:=\{a:(a,n)\in c_1\}$ and $m_2:=\{a:(a,n)\in c_2\}$,\\ then $\mid m_1\cup m_2\mid=\mid c_3\mid $.
\end{enumerate}
In $d_1$, $m+n$ refers to the usual sum of integers; in $d_4$, $n$ is a free variable.\\
With this operator, we define $\sum_{i=1}^n (c_i)$ as the successive application of $+^c$ to the elements of the sequence $(c)_{i=1}^n$.\\ As an example, $\sum_{i=1}^4 (c_i)=((c_1+^{c}c_2)+^{c}c_3)+^{c}c_4$. With this ``series", we define the function $\gamma$, being $\delta (S^{\pi})\subset\mathbb{U}$ the range of $\delta$,
\begin{equation*}
    \begin{array}{rl}
        \gamma:\delta(S^{\pi})\to & \pi(\mathbf{A})\\
        C\mapsto & D=\gamma(C)
    \end{array}
\end{equation*}
Where $D=\gamma(C)$ if and only if, being $(c)_{i=1}^n$ a sequence of all the elements of $C$, $D=\sum_{i=1}^n (c_i)$.\\
The last step consists in transforming $D$ back into a multiset, we do this with our last function $\pi^{-}$.
\begin{equation*}
    \begin{array}{rl}
        \pi^{-}:\pi(\mathbf{A})\to &\mathbf{A} \\
        D\mapsto & E=\pi^{-}(D)
    \end{array}
\end{equation*}
$\pi^{-}$ is defined as
$$ E=\pi^{-}(D) \iff E\sim D $$
Similar to $\pi$, this function is also injective.
Furthermore, $\pi^{-}$ is also surjective since its range is the whole codomain $\mathbf{A}$.\\
Finally, we resume all of these operations with the function $pre-\phi$,
\begin{equation*}
    \begin{array}{rl}
        pre-\phi:\mathbf{A}\to &\mathbf{A} \\
        A\mapsto & E=pre-\phi(A)
    \end{array}
\end{equation*}
$pre-\phi$ is defined as
$$ E=pre-\phi(A) \iff E=\pi^{-}(\gamma(\delta(\pi(A))))$$

\subsection{One NL-set per image of $pre-\phi$}\label{secA3}
Given $pre-\phi$ definition, we can prove that there is only one NL-set per image of $pre-\phi$. Let $\diamond^{'}$ be a relation between msets of ordered pairs and NL-sets as follows
$$\alpha\diamond^{'} \beta \iff (\alpha^{*}\subset \mathcal{P}(\beta\times\beta)\land(\forall x\in\beta \exists (y_1,y_2)\in\alpha:x=y_1\lor x=y_2)) $$
Let $\mathbf{E}$ be a non-empty collection of images of the function $pre-\phi$ and $\mathbf{L}$, an arbitrary non-empty set of NL-sets as previously, then  the following proposition is true.
\begin{proposition} \label{proponeNLperprephi}
$$ \forall A\in\mathbf{E}\forall X,Y\in\mathbf{L}, ((A\diamond^{'} X\land A\diamond^{'} Y)\implies X=Y) $$
\end{proposition}
\begin{proof}
Let $P$ be the proposition, then
$$\neg P\equiv \exists A\in\mathbf{E}\exists X,Y\in\mathbf{L},(A\diamond^{'} X\land A\diamond^{'} Y)\land X\neq Y $$
\begin{equation*}
\begin{array}{ccc}
\neg P & \implies & ((\exists x\in X:x\not\in Y)\lor (\exists y \in Y:y\not\in X) )\\
& \implies & ((\exists(\alpha_1,\alpha_2)\in A:\alpha_1\not\in Y\lor \alpha_2\not\in Y)\lor\\
&&((\exists(\beta_1,\beta_2)\in A:\beta_1\not\in X\lor \beta_2\not\in X))\\
& \implies & ((\exists(\alpha_1,\alpha_2)\in A: (\alpha_1,\alpha_2)\not\in A)\lor\\
&&(\exists(\beta_1,\beta_2)\in A: (\beta_1,\beta_2)\not\in A))\\
& \implies & A\neq A
\end{array}
\end{equation*}
\begin{flushright}
$(\Rightarrow\!\Leftarrow)$
\end{flushright}
\begin{center}
$\therefore \forall A\in\mathbf{E}\forall X,Y\in\mathbf{L}, ((A\diamond^{'} X\land (A\diamond^{'} Y)\implies X=Y) $
\end{center}
\end{proof}
\subsection{Lemma for unnested forms}
\label{lemma_case1}
\begin{lemma}
\label{lemma_1}
Let $\phi(a)=(X,pre-\phi(A_a))$ and $\phi(b)=(Y,pre-\phi(A_b))$ be multidigraphs obtained by applying $\phi$ to the $\Upsilon$-graphs $a$ and $b$, respectively. Let $P_a=(c_{i1})_{n\times 1}$ be a column matrix where every element $c_{i1}=Count_{pre-\phi(A_a)}(\alpha_i)$, $\alpha_i\in pre-\phi(A_a)^{*}$ with $i\in [1,n]\cap \mathbb{Z}$, being $n$ the cardinality of $pre-\phi(A_a)^{*}$. Analogously for $b$, let $P_b=(d_{j1})_{m\times 1}$ be a column matrix where every element $d_{j1}=Count_{pre-\phi(A_b)}(\beta_j)$, $\beta_j\in pre-\phi(A_b)^{*}$ with $j\in [1,m]\cap \mathbb{Z}$, being $m$ the cardinality of $pre-\phi(A_b)^{*}$. Then the following is true.
$$\forall a,b\in \mathbf{P}, \phi(a)\cong_\alpha \phi(b)\implies (\exists T : TP_a=P_b) $$
\end{lemma}
In other words, the elements of $P_a$ can be rearranged into $P_b$. This does not imply that $P_a=P_b$ although it certainly does not exclude it ($T=I_{n\times n}$).
\begin{proof}
The first thing to realize is that if $\phi(a)\cong_\alpha \phi(b)$, then there exists a bijection $f$ between $X$ and $Y$ such that
$$Count_{pre-\phi(A_a)}(p,q)=Count_{pre-\phi(A_b)}(f(p),f(q))$$
for any $p,q\in pre-\phi(A_a)$. Since for any $(p,q)\in pre-\phi(A_a)^{*}$ there exists a unique element $(f(p),f(q))\in pre-\phi(A_b)^{*}$ and viceversa, we can define a function $g$ which is also a bijection; however, this function will be defined from $pre-\phi(A_a)^{*}$ to $pre-\phi(A_b)^{*}$ which, in turn, implies that $pre-\phi(A_a)^{*}$ and $pre-\phi(A_a)^{*}$ have the same cardinality. With this, we can affirm that the column matrices $P_a$ and $P_b$ have the same number of elements, i.e., $n=m$.\\
Now, since every element $c_{i1}$ of $P_a$ is associated with one and only one element of $pre-\phi(A_a)^{*}$ and the same for the elements $d_{j1}$ of $P_b$ with $pre-\phi(A_b)^{*}$, then, if $P_a^{\star}$ and $P_b^{\star}$ are the sets that contain the elements of $P_a$ and $P_b$, respectively, we can define the following function
$$ 
\begin{array}{cccc}
h : & P_a^{\star} &\to& P_b^{\star}\\
 & c_{i1}&\mapsto & d_{j1}
\end{array}
$$
with $d_{j1}=h(c_{i1})$ if and only if $(c_{j1}=Count_{pre-\phi(A_a)}(\alpha)\implies d_{j1}=Count_{pre-\phi(A_b)}(g(\alpha)))$. Since $g$ is a bijection, it is clear that $h$ is also a bijection. This implies that every element $d_{j1}$ can be defined as $h(c_{i1})$, in other words, $P_a=(c_{i1})_{n\times 1}$ and $P_b=(h(c_{i1}))_{n\times 1}$.
Now, by the definition of $g$ and the fact that $\phi(a)\cong_\alpha \phi(b)$, we say that $\forall c_{i1}\in P_a^{\star}, h(c_{i1})=c_{i1}$. Together with the new definition of $P_b$, this implies that $P_a^{\star}= P_b^{\star}$. Therefore $P_b$ can only be equal to $TP_a$ where $T$ is a row exchange matrix. We complete this proof by providing such $T$.\\
Let $T=(\alpha_{ij})$, where
$$ \alpha_{ij}=
\begin{cases}
1, & c_{j1}=d_{i1}\\
0, & c_{j1}\neq d_{i1}
\end{cases}$$
where $c$ represents an element of $P_a$ and $d$, an element of $P_b$.
\end{proof}

\subsection{Proof of proposition \ref{prop_sum}}
\label{proof_1}

Let $\alpha\in pre-\phi(A_x)$ be any vertex of $\phi(x)$ (i.e., an ordered pair), if $s_1,s_2,...,s_k$ are all synaptic configurations (elements of $A_x$) that include $\alpha$ ($\alpha\in s_i$ for all $i\in[1,k]\cap \mathbb{Z}$), then

$Count_{pre-\phi(A_x)}(\alpha)=Count_{A_x}(s_1)+ Count_{A_x}(s_2)+...+Count_{A_x}(s_k)$
\begin{proof}
    From the definition of $pre-\phi$ (appendix \ref{secA2}), $(\alpha,Count_{pre-\phi(A_x)}(\alpha))$ is an element of $D=\gamma(\delta(\pi(A_x)))$. However, if $s_1,...,s_k$ are the representations of the synaptic configurations that include $\alpha$, then
$$ (\alpha,Count_{pre-\phi(A_x)}(\alpha))=(\alpha,Count_{A_x}(s_1))+^{c}...+^{c}(\alpha,Count_{A_x}(s_k)) $$
since condition $d_1$ of the definition of $+^{c}$ (defined in \ref{plus_c_def}). In other words, $Count_{pre-\phi(A_x)}(\alpha)=Count_{A_x}(s_1)+...+Count_{A_x}(s_k)$.
\end{proof}
\subsection{Matrix for proposition \ref{prop_case4}}\label{secA4}
\begin{center}

$A=$
$\begin{bmatrix}
    0&0&0&0&0&0&0&0&0&0&0&0&0&0&0&0&0&0&1&0&1&0\\
    0&0&0&0&0&0&0&0&0&0&0&0&0&0&0&0&0&0&0&1&1&0\\
    0&0&0&0&0&0&0&0&0&0&0&0&0&0&0&0&0&1&0&0&0&1\\
    0&0&0&0&0&0&0&0&0&0&0&0&0&0&0&0&0&1&1&0&0&0\\
    0&0&0&0&0&0&0&0&0&0&0&0&0&0&0&0&1&0&0&1&0&0\\
    0&0&0&0&0&0&0&0&0&0&0&0&0&0&0&0&1&0&0&0&0&1\\
    0&0&0&0&0&0&0&0&0&0&0&0&0&0&0&1&0&0&0&1&0&0\\
    0&0&0&0&0&0&0&0&0&0&0&0&0&0&1&0&0&0&1&0&0&0\\
    0&0&0&0&0&0&0&0&0&0&0&0&0&1&0&0&0&1&0&0&0&0\\
    0&0&0&0&0&0&0&0&0&0&0&0&0&1&1&0&0&0&0&0&0&0\\
    0&0&0&0&0&0&0&0&0&0&0&0&1&0&0&1&0&0&0&0&0&0\\
    0&0&0&0&0&0&0&0&0&0&0&0&1&0&0&0&1&0&0&0&0&0\\
    0&0&0&0&0&0&0&0&0&0&1&1&0&0&0&0&0&0&0&0&0&0\\
    0&0&0&0&0&0&0&0&0&1&0&1&0&0&0&0&0&0&0&0&0&0\\
    0&0&0&0&0&0&0&1&0&0&0&0&0&0&0&1&0&0&0&0&0&0\\
    0&0&0&0&0&0&0&1&0&1&0&0&0&0&0&0&0&0&0&0&0&0\\
    0&0&0&0&0&0&0&0&1&0&0&0&0&0&1&0&0&0&0&0&0&0\\
    0&0&0&0&0&0&0&0&1&0&1&0&0&0&0&0&0&0&0&0&0&0\\
    0&0&0&0&0&0&1&0&0&0&0&0&0&1&0&0&0&0&0&0&0&0\\
    0&0&0&0&0&1&0&0&0&0&0&0&1&0&0&0&0&0&0&0&0&0\\
    0&0&0&0&1&0&1&0&0&0&0&0&0&0&0&0&0&0&0&0&0&0\\
    0&0&0&0&1&0&0&0&0&0&1&0&0&0&0&0&0&0&0&0&0&0\\
    0&0&0&1&0&1&0&0&0&0&0&0&0&0&0&0&0&0&0&0&0&0\\
    0&0&0&1&0&0&0&0&0&1&0&0&0&0&0&0&0&0&0&0&0&0\\
    0&1&0&0&1&0&0&0&0&0&0&0&0&0&0&0&0&0&0&0&0&0\\
    0&1&0&1&0&0&0&0&0&0&0&0&0&0&0&0&0&0&0&0&0&0\\
    1&1&0&0&0&0&0&0&0&0&0&0&0&0&0&0&0&0&0&0&0&0\\
    1&0&1&0&0&0&0&0&0&0&0&0&0&0&0&0&0&0&0&0&0&0\\
    0&0&1&0&0&0&0&0&1&0&0&0&0&0&0&0&0&0&0&0&0&0\\
    0&0&1&0&0&0&0&1&0&0&0&0&0&0&0&0&0&0&0&0&0&0\\
\end{bmatrix}
$
\end{center}





\printbibliography


\end{document}